# Mariner 2 and its Legacy: 50 Years On

Jeremy Bailey

*School of Physics, University of New South Wales, NSW, 2052, Australia*

**Summary:** Fifty years ago, NASA's Jet Propulsion Laboratory (JPL) built and flew the first successful spacecraft to another planet: Mariner 2 to Venus. This paper discusses the context of this mission at a crucial phase in the space race between the USA and the USSR and its results and legacy. As its first major success, Mariner 2, helped to cement JPL's position as a centre for robotic planetary exploration. Mariner 2 successfully solved the scientific problem of the high temperature observed for Venus by ground-based radio telescopes. It also pioneered new techniques for observing the atmosphere of a planet from space, which were subsequently developed into the microwave sounding and infrared sounding techniques for observing the Earth atmosphere. Today these techniques provide some of the most important data for constraining weather forecasting models, as well as a key series of data on the Earth's changing climate.

## Introduction

On August 27$^{th}$ 1962, the Mariner 2 spacecraft built by NASA's Jet Propulsion Laboratory (JPL) was successfully launched on its way to Venus. On December 14$^{th}$ 1962 it passed within 35,000 km of Venus and observed the planet with its onboard instruments and returned the resulting data to Earth [1]. It was the first successful spacecraft to another planet.

The Mariner 2 mission came at the height of the space race between the USA and USSR, and at a time when the USSR was making all the running. The Soviet Union had shocked the world with the launch of Sputnik 1 in 1957. The launch was made possible by the development of the world's first Intercontinental Ballistic Missile (ICBM), the R-7, by the team led by Sergei Korolev in Moscow. A month after the first successful test flight of the R-7 it was used to put Sputnik 1 in orbit [2]. By 1959 Korolev had added an additional upper stage to his R-7 to make the 8K72 launch vehicle used to launch a series of spacecraft to the Moon. Luna 1 flew past the Moon in January 1959. Luna 2 impacted on the Moon's surface in September 1959, and Luna 3 flew round the Moon and photographed its far side in October 1959 [3]. The same launch vehicle was used to launch the 4.7 tonne Vostok manned spacecraft that carried Yuri Gagarin into orbit in April 1961 [4].

Over the same period the USA was playing catch up. Early US space launches, such as the Explorer and Vanguard satellites, and Pioneer lunar and interplanetary spacecraft were smaller spacecraft launched by smaller rockets, and the US was repeatedly beaten by the Soviet Union to key space "firsts". In this context the achievement of the first successful planetary mission with Mariner 2 in 1962 came as something of a surprise.

By 1960 Korolev's team had added two upper stages to the R-7 to make the 8K78 (later known as the Molniya) and was using this to attempt launches to the planets. Two Mars launch attempts were made in October 1960 but both were lost when the third stage of the launch vehicle failed to ignite [5]. Then two Venus launches were attempted in February

1961. The first failed to leave Earth orbit, but the second (later designated Venera 1) was successfully placed on its path to Venus. It did not last long. After a few days the spacecraft's attitude stabilization and thermal control failed and radio contact was lost. Venera 1 did fly within 100,000 km of Venus, but attempts to receive signals from it (even using the powerful radio telescope at Jodrell Bank in England) were unsuccessful [6].

## JPL and Mariner

The Jet Propulsion Laboratory grew out of the work of a group of rocket enthusiasts at CalTech who began experimenting with rocket engines in 1936. During World War II the group developed JATO units, rockets used to assist aircraft take-off. JPL was formally established in 1944 [7]. After the war JPL, worked on the development of missiles including the liquid fuelled Corporal [8] and solid fuelled Sergeant [9]. In 1954, William Pickering became director of JPL and remained in the role until 1976 [10].

The first satellite launched by the US, Explorer 1, was the result of collaboration between JPL and the Army Ballistic Missile Agency (ABMA). The launch vehicle (Juno 1) was based on ABMA's Redstone missile developed by Wernher von Braun. The upper stages and the satellite were built by JPL, with the upper propulsive stages being made from a cluster of miniature "Baby Sergeant" solid propellant rockets [11]. The satellite carried a radiation detector built by Iowa State University's James Van Allen, and the measurements from this and subsequent Explorer missions led to the discovery of the Van Allen radiation belts [12].

Following the success of Explorer, JPL became part of the new National Aeronautics and Space Administration (NASA) formed in 1958 to coordinate the USA's civilian space program. By the end of 1959 the role of JPL was determined to be the development of unmanned spacecraft to the Moon and planets. The lunar missions would be known as Rangers, and spacecraft to Mars and Venus would be given the name Mariner [13].

A major issue for the US lunar and planetary missions was to find a suitable launch vehicle. While the USSR used the R-7 ICBM and derivatives for all space launches, most early US launch attempts were based on smaller rockets. America's own ICBM, the Atlas, went through test flights in 1958 and 1959 [14], but Atlas was smaller than the R-7 and needed additional upper stages for lunar and planetary launches. The first configuration tried was the Atlas-Able, which added the upper stages of the Vanguard launch vehicle to the Atlas. Four Pioneer launch attempts with this launch vehicle in 1959 and 1960 all resulted in failure [15].

JPL and NASA had a proposal for a new vehicle called the Atlas-Vega, but this project was dropped when it was realized that the Air Force's Agena B could be used as an Atlas upper stage to provide similar performance [16]. The other project under development was the Centaur [17], a high performance liquid hydrogen fuelled upper stage designed specifically for use with the Atlas. The Atlas-Centaur combination was the most powerful available, and JPL designed its first Mariner spacecraft concept named Mariner-A for an Atlas-Centaur launch to Venus in the 1962 launch window. Mariner-A would have been a spacecraft of around 500 kg [18].

By August 1961 it became clear that the Centaur would not be ready for the planned 1962 launches. Instead a concept for a spacecraft called Mariner-R, based partly on the Ranger spacecraft being designed for lunar missions, was rapidly developed. This would have a mass of only ~200 kg allowing it to be launched with the smaller Atlas-Agena-B. According to Mariner project manager Jack James, "In a three week period from 8 August 1961 to 1 September 1961, a conceptual design was completed" [19]. That left just nine months for the spacecraft components to be designed, fabricated, tested and assembled, before three Mariner-R spacecraft (two flight models and a spare) were shipped to the launch site at Cape Canaveral in the first week of June 1962 [19].

The Mariner-R spacecraft carried six science instruments as shown in table 1 [20].

| Instrument | Mass (kg) | Max Power (W) | Investigators |
|---|---|---|---|
| Microwave Radiometer | 9.9 | 8.9 | A. H. Barrett (MIT) <br> J. Copeland (Army Ordnance Missile Command) <br> D. E. Jones (JPL) <br> A. E. Lilley (Harvard Observatory) |
| Infrared Radiometer | 1.2 | 2.1 | L. D. Kaplan (JPL & Univ. Nevada) <br> G. Neugebauer (JPL) <br> C. Sagan (UC Berkeley) |
| Magnetometer | 2.1 | 6.0 | P.J. Coleman (NASA) <br> L. Davis (CalTech) <br> E.J. Smith (JPL) <br> C.P. Sonnett (NASA) |
| Radiation Detectors | 1.3 | 0.4 | H.R. Anderson (JPL) <br> H.V. Neher (CalTech) <br> J. A. Van Allen (State University of Iowa) |
| Cosmic Dust Detector | 0.8 | 0.08 | W. M. Alexander (GSFC) |
| Solar Plasma Analyser | 2.2 | 1.0 | M. Neugebauer (JPL) <br> C.W. Snyder (JPL) |

*Table 1 — Scientific Instruments for the Mariner-R Spacecraft (from ref 20).*

It is clear from Table 1 that the Microwave Radiometer stands out with a mass greater than all the other instruments combined. This was, indeed, the key instrument on Mariner designed to solve an intriguing problem about Venus.

## Radiometry of Venus

A radiometer is an instrument that measures thermal radiation. Since this radiation depends on the temperature of the source according to the black body radiation laws, it can be used to measure temperature remotely. In the 1920's a few pioneering scientists used radiometry to measure the temperature of other objects in the solar system. Among them were Edison Pettit and Seth Nicholson who worked at the Mount Wilson Observatory in California. They used vacuum thermocouples that they built themselves [21], placed at the focus of Mount Wilson's 100-inch Hooker telescope, then the largest telescope in the World. The instruments were sensitive to infrared radiation at wavelengths of about 10 μm.

In 1924 Pettit and Nicholson reported preliminary measurements of the temperature of Venus. They measured temperatures "near the freezing point" [22]. It wasn't until 31 years later, in 1955 [23], that they published a detailed analysis of their 1924 observations giving temperatures of about 230-240 K. Similar temperatures were measured by Sinton and Strong [24]. These measurements apply to the top of the cloud layer that we now know is at an altitude of around 70km. To see what was going on below the clouds would require a different approach.

The microwave radiometer was invented in 1942, and was a by-product of World War 2 radar research. Microwave radar had been made possible by the British invention of the cavity magnetron [25]. This secret development was shared with the USA and led to the establishment of the MIT Radiation Laboratory (RadLab) to further develop microwave radar [26]. It was here that Robert Dicke built the first microwave radiometer, a radio receiver sensitive enough to be able to detect thermal radiation [27].

After the war many of the scientists who had worked on wartime radar began to explore the new field of radio astronomy. While much of the early work on radio astronomy involved observations at metre wavelengths a few researchers explored the shorter microwave wavelengths. At the Naval Research Laboratory in Washington DC, Cornell Mayer and Fred Haddock built a 50-foot microwave telescope mounted on a gun mount on the roof of one of the NRL buildings [28]. Using this telescope, in 1956, Mayer made measurements of the temperature of Venus using a Dicke type radiometer working at a wavelength of 3.15 cm. He obtained temperatures ranging from 560 to 620 K [29]. Other observations soon followed and confirmed the high temperatures [30]. But what was being measured?

According to Carl Sagan [31] the high temperatures measured by the microwave instruments were thermal emission from the surface or lower atmosphere heated by a very efficient greenhouse effect. The key difference between the microwave and infrared measurements is that microwave radiation is unaffected by the clouds and so can measure the true surface temperature, whereas infrared radiation cannot see through the clouds and measures only the cloud tops. However, many scientists were reluctant to accept that the greenhouse effect was sufficient to heat the surface of Venus to such extreme temperatures. Other mechanisms were suggested in which the microwave radiation originated in the ionosphere of Venus [31].

The microwave radiometer on Mariner 2 was designed to carry out observations that would clearly distinguish these two models. By scanning across the disk of Venus it would determine if the microwave radiation increased or decreased at the limb of the planet. Limb darkening would indicate that Sagan's greenhouse model was correct because radiation coming from the surface would suffer more absorption on the way out near the limb. Limb brightening would favour the ionosphere model since a greater path length of the ionosphere would be observed near the limb. This test could not be carried out using earth-based radio telescopes as their spatial resolution, at that time, was far too small.

## Voyage to Venus

Mariner 1, the first of the Mariner-R spacecraft was launched on July 21 1962. Five minutes into the flight the Atlas-Agena had to be destroyed after a guidance system failure. Mariner 1 continued to transmit until it hit the Atlantic ocean 357 seconds after lift-off [16]. For JPL it was all too familiar. The first four of their Ranger spacecraft to the Moon had also resulted in failure.

Another Atlas-Agena was readied to launch Mariner 2. After some delays the launch finally occurred on August 27 1962. This time the Atlas-Agena was successful. The Agena-Mariner combination was placed in a parking orbit and then the Agena fired again to launch Mariner 2 on its way to Venus.

The heart of Mariner 2 was a system called the "Central Computer and Sequencer" (CC&S). This was not a computer as we would understand the term today. The CC&S was basically a timer that counted the time since launch and initiated operations and manoeuvres at the appropriate times [32]. Following launch the CC&S instructed the spacecraft to deploy its solar panels and turn on its attitude control system, which pointed the spacecraft towards the Sun and the high-gain antenna at the Earth.

On September 4 the CC&S carried out the mid-course manoeuvre, which involved a motor-burn to adjust the spacecraft's velocity to put it on course for its encounter with Venus. The required burn duration and direction parameters were provided by ground command, based on analysis of the spacecraft's orbit to that point.

Back on Earth the Soviet team were still trying to launch their planetary spacecraft. Three Venus launches in the same launch window as Mariner 2 all failed. However, on November 1st 1962 Mars 1 was successfully launched on a trajectory to Mars [33]. Should Mariner 2 fail, the Soviet Union could still achieve the first successful flight to another planet.

And all was not well with Mariner 2. A short circuit occurred in one of the two solar panels, the fault fixed itself a few days later and then recurred. Fortunately as Mariner was approaching the Sun the power from a single solar panel would be sufficient, although the scientific instruments had to be turned off for 8 days to conserve power [16].

As well as the solar panel problem, Mariner 2 was overheating. The thermal control system designed to keep the spacecraft at a safe operating temperature was not working well enough. On November 16 one temperature sensor reached its limit [34] and temperatures were generally 22 °C higher than expected [16]. Several of the spacecraft systems were operating at or above their maximum temperatures, including the crucial earth-sensor that kept Mariner 2 correctly oriented, with its antenna pointed at Earth. Failures could be expected at any time. According to project manager Jack James, "Now we could only wait and hope" [16].

On December 9, just 5 days before encounter, four of the spacecraft telemetry measurements that were regularly returned to Earth failed. On December 12 the CC&S failed to issue a regular update pulse that should occur every 1000 minutes [34]. This indicated that the CC&S might fail to initiate the encounter activities and so a ground command was sent to initiate the encounter sequence with Venus.

On December 14 the overheated Mariner 2 flew past Venus at a distance of 35,000 km with 7 of its 18 temperature sensors now at their limit [34]. However, everything worked as expected. The radiometers performed their scans across the disk of Venus and returned the crucial data on the planet's temperature. All the science instruments functioned properly. The spacecraft continued to operate until January 3 1963.

## Results

The Mariner 2 microwave radiometer experiment confirmed the high temperatures observed in ground-based observations. Results from three scans across the disk showed limb darkening and thus showed conclusively that the surface of Venus was hot [16, 35, 36] as had been suggested by Sagan, and ruled out the ionosphere model. The results would lead to the runaway greenhouse model for the evolution of the Venus atmosphere [37].

Mariner 2's infrared radiometer confirmed the cold cloud-top temperatures (220-230 K) obtained by the ground-based measurements and showed no evidence of breaks in the cloud coverage that could permit deeper views into the atmosphere [38].

Mariner 2's other instruments such as the magnetometer and radiation detectors made valuable studies of the magnetic field and radiation environment in interplanetary space [1]. During the encounter with Venus, however, no change was detected in the magnetic field and radiation levels. This indicated that Venus was very different to Earth with a much weaker magnetic field and hence no extended magnetosphere.

From precise tracking of Mariner's orbit as it passed Venus it was possible to derive a much improved value for the mass of Venus [16].

## From Radiometers to Sounders

If microwave and infrared radiometers could make valuable observations of the temperature structure in the Venus atmosphere, then could the same techniques play a role in studying the Earth's weather and climate? This was what several of the Mariner scientists began to investigate in the years following Mariner 2's mission.

Alan Barrett, a radio astronomer at MIT, and an investigator on the Mariner 2 microwave radiometer experiment, began to develop the concept for a microwave "sounder" that would observe the temperature structure of the atmosphere from space. The way to make a sounder was to make use of an absorption band, in this case one due to oxygen at around 60 GHz. By using microwave radiometers at different frequencies within this band, with different amounts of absorption, temperatures could be measured at different heights in the atmosphere.

Barrett had a number of his graduate students work on the concept. One of these, William B. Lenoir (who would go on to be a NASA astronaut) tested the concept using a balloon-borne radiometer [39]. Others used ground and airborne instruments, but the real aim was to put an instrument on a satellite that could look down on Earth and make global 3D maps of the temperature distribution in the atmosphere.

The opportunity came in 1972 with the Nimbus 5 satellite and an instrument called NEMS (Nimbus-E Microwave Spectrometer). Another former Barrett student, David Staelin, then on the MIT faculty, was the principal investigator for the instrument, with Barrett as a co-investigator. The instrument successfully demonstrated the concept of microwave temperature sounding from space [40].

The follow up to this experimental instrument was the microwave sounding unit (MSU), a four channel microwave sounder [41] that was installed as standard on a series of meteorological satellites starting with TIROS-N in 1978 and ending with NOAA-14 in 1994. It was then superseded by an improved instrument, the Advanced Microwave Sounding Unit (AMSU), with 15 channels. AMSU is currently installed on a number of NOAA satellites, on NASA's Aqua and on the European MetOp satellites.

In parallel with the microwave development, Lewis Kaplan, a member of the Mariner 2 infrared radiometer team was making similar proposals for an infrared sounder. In fact even before Mariner he had proposed atmospheric temperature sounding using a $CO_2$ band at 15 µm [42]. The first satellite instrument to use this technique was the Nimbus 3 Satellite Infrared Spectrometer (SIRS) launched in 1969. The present versions of such instruments are the Atmospheric Infrared Sounder (AIRS) on the NASA Aqua satellite and the Infrared Atmospheric Sounding Interferometer (IASI) on the European MetOp satellites.

Weather forecasting today is based on sophisticated computer models that "assimilate" data from a wide variety of sources including ground weather stations and remote sensing data from satellites. It is possible to analyse these systems to determine which observations contribute most to the accuracy of the forecast. Such an analysis carried out at the European Centre for Medium-Range Weather Forecasting (ECMWF) [43] shows that the instruments at the top of the list are AMSU, AIRS and IASI, the very instruments that can trace their heritage to Mariner 2.

The more than 30 year record of data from microwave sounding using MSU and AMSU also provides an important dataset on the earth's warming atmosphere [44], showing the trend in temperature rise, as well as the geographical distribution, with global warming being greatest in the arctic region.

## Conclusions

Mariner 2 is an important mission not just because of the obvious milestone of being the first successful spacecraft to another planet. Its success in 1962 marks the beginning of the turnaround of fortunes in the space race. As JPL's first successful lunar or planetary mission it helped to secure JPL's future and was the forerunner of a long record of success in planetary exploration.

It also did important science establishing the reality of the extreme greenhouse effect in the atmosphere of Venus. And the radiometers that flew on it are the predecessors of the infrared and microwave sounders that circle the earth today and are major contributors to the accuracy of the weather forecasts we use every day.